\documentclass{kluwer}    % Specifies the document style.
\newdisplay{guess}{Conjecture}
\usepackage{epsfig}

\begin{document}                                                                                   
\def\lsim{\lower 2pt \hbox{$\, \buildrel {\scriptstyle <}\over
         {\scriptstyle \sim}\,$}}

\begin{article}
\begin{opening}         
\title{Pulsar Wind Nebulae in EGRET Error Boxes} 
\author{Mallory S.E. \surname{Roberts} \email{roberts@physics.mcgill.ca}}  
\institute{McGill University / Eureka Scientific}
\author{Crystal L. \surname{Brogan}}
\author{Bryan M. \surname{Gaensler}}
\author{Jason W.T.  \surname{Hessels}}
\institute{Institute for Astronomy, Harvard-Smithsonian CfA, and McGill U.}
\author{C.-Y. \surname{Ng}}
\author{Roger W. \surname{Romani}}
\institute{Stanford University}
\runningauthor{Roberts et al.}
\runningtitle{PWN in EGRET Error Boxes}
\date{June 2, 2004}

\begin{abstract}
A remarkable number of pulsar wind nebulae (PWN) are coincident with 
$EGRET$ $\gamma$-ray sources.  X-ray and radio imaging studies of unidentified 
$EGRET$ sources have resulted in the discovery of at least 6 new pulsar wind 
nebulae (PWN). Stationary PWN  (SPWN) appear to be associated with 
steady $EGRET$ sources with hard spectra, typical for $\gamma$-ray pulsars. Their 
toroidal morphologies can help determine the geometry of the pulsar which is 
useful for constraining models of pulsed $\gamma$-ray emission. Rapidly moving PWN 
(RPWN) with more cometary morphologies seem to be associated with variable 
$EGRET$ sources in regions where the ambient medium is dense compared to what is typical for the ISM.
\end{abstract}

\end{opening}           

\vspace{-0.3cm}
\section{Coincidence of PWN with EGRET Sources}  

\vspace{-0.3cm}
A pulsar wind nebula (PWN) is a X-ray, radio, and/or H$\alpha$ nebula 
caused by a relativistic particle wind from an energetic pulsar (see 
\inlinecite{krh04} and \inlinecite{gae04} for recent reviews). Of the $\sim 50$ known PWN, a large fraction
are coincident with $EGRET$ sources. If we consider only the $\sim 30$ sources
with significant emission above 1 GeV (listed in the catalog of \inlinecite{lm97}) which are probably associated with our 
Galaxy, nearly half the error 
ellipses contain a known PWN, and several of the others have not yet been 
searched carefully for X-ray nebulae.  
In Table~\ref{pwntab}, we list all the known PWN coincident with $EGRET$ sources
(for references 
to these sources, see http://www.physics.mcgill.ca/$\sim$pulsar/pwncat.html).
We also list the morphological type (S or R, see below) if 
it can be determined, the pulsar spin-down energy $\dot E$ if pulsations have
been detected, and the variability index $V_{12}$ of the $EGRET$ source
as defined and determined by \inlinecite{ntgm03}. $10^{-V_{12}}$ is the integrated
liklihood that the variability measure $\tau \equiv \sigma_{F}/<F>\, \le 0.12$, 
or the $> 100$ MeV flux is consistent with being constant. In other words, 
$V_{12} > 1$ indicates a $> 90\%$ confidence level that the $\gamma$-ray source 
is variable. No value indicates there is no evidence for variability. 
\begin{table*}[h!]
\caption[]{Pulsar Wind Nebulae Coincident with $EGRET$ Sources}
\label{pwntab}
\begin{tabular}{lccccc}
\hline
Name & 3EG Name & GeV Name & Type & log$\dot E$ & $V_{12}$ \\ 
\hline
CTA 1 & 3EG J0010+7309 & GeV J0008+7304 & ? & -- & 0.40 \\
Crab & 3EG J0534+2200 & GeV J0534+2159 & S & 38.7 & -- \\
Geminga & 3EG J0633+1751 & GeV J0634+1746 & R & 34.5 & 0.16 \\
Vela & 3EG J0834$-$4511 & GeV J0835$-$4512 & S & 36.8 & 0.61 \\
MSH 11-6$2$ & 3EG J1102$-$6103 & -- & R & -- & -- \\
PSR J1016$-$5857 & 3EG J1013$-$5915 & -- & ? & 36.4 & 0.18 \\
PSR B1046$-$58 & 3EG J1048$-$5840 & GeV J1047$-$5840 & ? & 36.3 & -- \\
PSR J1420$-$6048 & 3EG J1420$-$6038 & GeV J1417$-$6100 & ? & 37.0 & 1.59 \\
Rabbit & 3EG J1420$-$6038 & GeV J1417$-$6100 & R & -- & 1.59 \\
PSR B1706$-$44 & 3EG J1710$-$4439 & GeV J1709$-$4430 & S & 36.5 & -- \\ 
G359.89$-$0.08 & 3EG J1746$-$2851 & GeV J1746$-$2854 & R & -- & 2.35 \\
G7.4$-$2.0 & 3EG J1809$-$2328 & GeV J1809$-$2327 & R & -- & 3.93 \\
G18.5$-$0.4 & 3EG J1826$-$1302 & GeV J1825$-$1310 & R & -- & 3.22 \\
PSR B1853+01 & 3EG J1856+0114 & GeV J1856+0115 & R & 35.6 & 1.57 \\
3C 396 & 3EG J1903+0550 & -- & ? & -- & 0.42 \\
CTB 87 & 3EG J2016+3657 & -- & ? & -- & 0.63 \\
PSR J2021+3651 & 3EG J2021+3716 & GeV J2020+3658 & S & 36.5 & 0.71 \\
PSR J2229+6114 & 3EG J2227+6122 & GeV J2227+6101 & S & 37.4 & 0.21 \\
\end{tabular}
\vspace{-1cm}
\end{table*}

At least six of these PWN were discovered through X-ray searches of the 
$EGRET$ error box, indicating that whatever the physical association,
$\gamma-$ray sources in the Galactic plane can serve as guides to the location
of energetic pulsars that otherwise might be difficult to find. 

\vspace{-0.3cm}
\section{Pulsar Geometry From Stationary Pulsar Wind Nebulae}

\vspace{-0.3cm}
The environment of a young pulsar is the expanding ejecta of its progenitor
star. This allows for fast expansion of the young PWN relative to the
kick velocity imparted to the pulsar. Therefore, the motion of the pulsar
does not greatly affect the structure of the PWN. The underlying 
symmettries of the pulsar's spinning magnetic field are then reflected
in the structure of the X-ray emitting PWN. We refer to these as stationary PWN (SPWN). SPWN tend to have some sort of equatorial toroidal structure and 
polar jets. The apparent ellipticity and brightness variations 
of these toroidal structures can be used to infer the orientation
of the pulsar's spin axis relative to the observer. In addition, the magnetic
inclination angle may relate to the thickness of the torus in the
case of single or concentric torii such as the Crab, or the separation between
two parallel torii such as is apparently the case for the Vela PWN.
For details in the methodology used to model observed SPWN torii, 
see \inlinecite{nr04}. 

This geometrical information is crucial for interpreting $\gamma$-ray pulse
profiles and hence understanding the particle acceleration mechanisms in 
pulsar magnetospheres. It is therefore important to infer the geometry 
of many torii around $\gamma$-ray emitting pulsars. Unfortunately, in 
only a handful of cases is the PWN structure clear enough 
for unambiguous interpretation of the geometry, even with the 
high resolution of the Chandra X-ray observatory. Only three of the known
$\gamma$-ray pulsars (Crab, Vela, and PSR B1706$-$44) have torii amenable to
fitting. It is therefore important to expand the number of potential 
$\gamma$-ray emitting pulsars with torii. X-ray imaging of unidentified
$EGRET$ source error boxes has resulted in the detection of two 
energetic young pulsars with well structured nebulae: PSR J2229+6114 
\cite{hcg+01} and PSR J2021+3651 \cite{hrr+04}. 
\begin{figure}[h!] 
\vspace{-0.5cm}
\centering
\epsfig{file=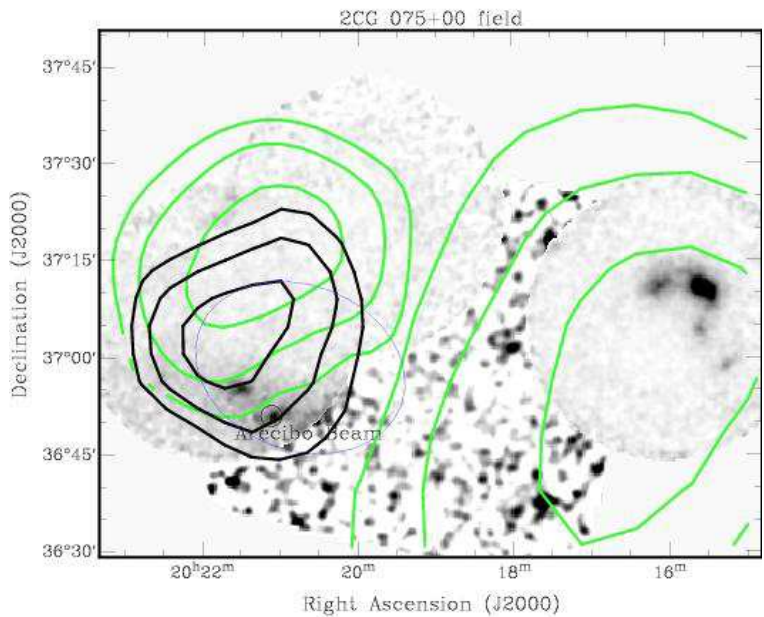, width=10cm}
\caption{X-ray image of the region of the COS B 
source 2CG 075+00 containing two 3rd EGRET Catalog
sources (light green contours), one GeV Catalog source (ellipse) and the refit 
GeV contours. The Arecibo beam pointing is shown which was used to discover
PSR J2021+3651.}
\label{2021_find}
\end{figure}

\begin{figure}[h!] 
\centering
\epsfig{file=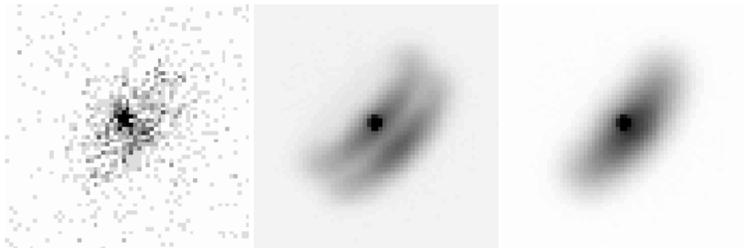, width=10cm}
\caption{Chandra image of PSR J2021+3651 with best fit double and single 
torus models (Hessels et al. 2004).}
\label{2021_torus}
\end{figure}

The methodology used to discover PSR J2021+3651 is illustrative of how to 
discover and measure usable torii. The $\gamma$-ray source is in the 
crowded Cygnus region, which
contains two highly significant GeV catalog sources which 
are not listed in the 3rd $EGRET$ Catalog.
A refitting of the region using only photons above 1 GeV but including
all known $EGRET$ sources was performed in order to obtain the best position
(see Fig.~\ref{2021_find}). 
A low resolution, wide field, hard X-ray observation was made using the 
$ASCA$ GIS detectors, discovering an apparent point source at the 
$\sim 1^{\prime}$ resolution of $ASCA$ \cite{rrk01}. 
This allowed the $~3^{\prime}$ beam
of the Arecibo telescope to perform a very deep search of the source to 
discover the $\sim 0.1$~mJy pulsar with spin properties virtually
identical to PSR B1706$-$44 \cite{rhr+02}. A 20~ks Chandra $ACIS$ image 
of the X-ray source revealed a compact nebula with a morphology 
highly suggestive of a double torus with polar jets. However, the
scarcity of counts in the image leaves some ambiguity as to whether or
not it is a single or double torus, and whether the jets are real
(Fig.~\ref{2021_torus}). 

If the double torus interpretation is made, then the separation of the 
two torii can be assumed to be an indication of the magnetic inclination
angle. This can be sanity checked by looking at the intrinsic radio pulse 
width, which may indicate whether the pulse is predominantly a core or conal 
component, and by measuring the radio polarization sweep and fitting it to 
a \inlinecite{rc69a} model.

Unfortunately, the above 5 sources are the only known pulsars associated with 
$EGRET$ sources whose nebulae are large enough and bright enough to 
be useful for fitting. However, there are several
more PWN with toroidal morphologies which either do not yet have detected 
pulsations
or are not yet associated with a $\gamma$-ray source. Even with $GLAST$,
it is likely that only $\sim 10$ $\gamma$-ray pulsars will be found 
whose geometry can be well constrained by their nebulae with Chandra imaging.  
We can only hope that will be enough to determine the location in the 
magnetosphere where $\gamma$-rays are emitted.

\vspace{-0.3cm}
\section{Rapidly Moving Pulsar Wind Nebulae Associated With Variable $EGRET$ Sources}

\vspace{-0.3cm}
The SPWN discussed above tend to be associated with hard spectrum 
($\Gamma \lsim 2$), steady ($V_{12} < 1$) $EGRET$ sources. Such sources 
tend to be nearly 100\% pulsed in $100-10000$ MeV $\gamma$-rays, 
and be the youngest and most energetic pulsars. After a few thousand years, 
the reverse shock of the supernova blast wave reaches the PWN, after which
the environment is no longer the expanding ejecta causing the PWN 
expansion to slow dramatically and become subsonic. If the pulsar
has a significant space velocity, it will start to 
catch up to the forward edge of the PWN, drastically affecting the morphology
as ram-pressure becomes more important.
If the space velocity exceeds the local sound speed, then the PWN will 
be confined in the forward direction by a bow-shock. These cometary
shaped nebulae we refer to as rapidly-moving or ram-pressure confined
PWN (RPWN). In X-rays, they generally have a narrow outflow trailing the 
pulsar, with the pulsar itself typically having a spin-down energy 
such that log$\dot E \sim 35.5 - 36.5$ erg/s, somewhat less than the 
typical SPWN. This is expected given their on average greater age.    

\begin{figure}[h!] 
\centering
\epsfig{file=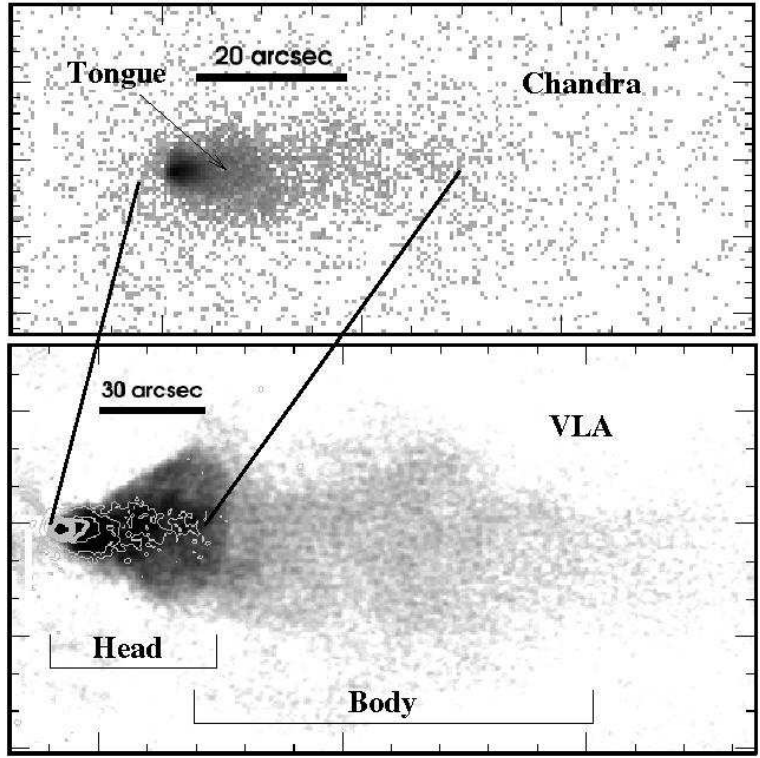, width=10cm}
\caption{Chandra X-ray and VLA radio images of the Mouse nebula showing the 
various structures (Gaensler et al. 2003).}
\label{mouse_img}
\end{figure}

Perhaps the best example of a RPWN is the Mouse nebula near the Galactic center.
The X-ray nebula of the Mouse is significantly brighter than other RPWN, 
and consequently shows significantly more structure (Fig.~\ref{mouse_img}).
A ``tongue" of brighter
emission is surrounded by a fainter halo, which is contained within the bright 
radio ``head" of the Mouse. A fainter radio body extends back from the head,
which eventually narrows into an extremely long tail \cite{gvc+03}. Hydrodynamic
models suggest the tongue is associated with the wind termination shock. 
However, much of the structure, especially downstream of the head, is
currently not understood. The generally short Chandra observations of 
other RPWN contain very few counts, and only the structure
corresponding to the Mouse's ``tongue" is clearly delineated. 

\begin{figure}[h!] 
\centering
\epsfig{file=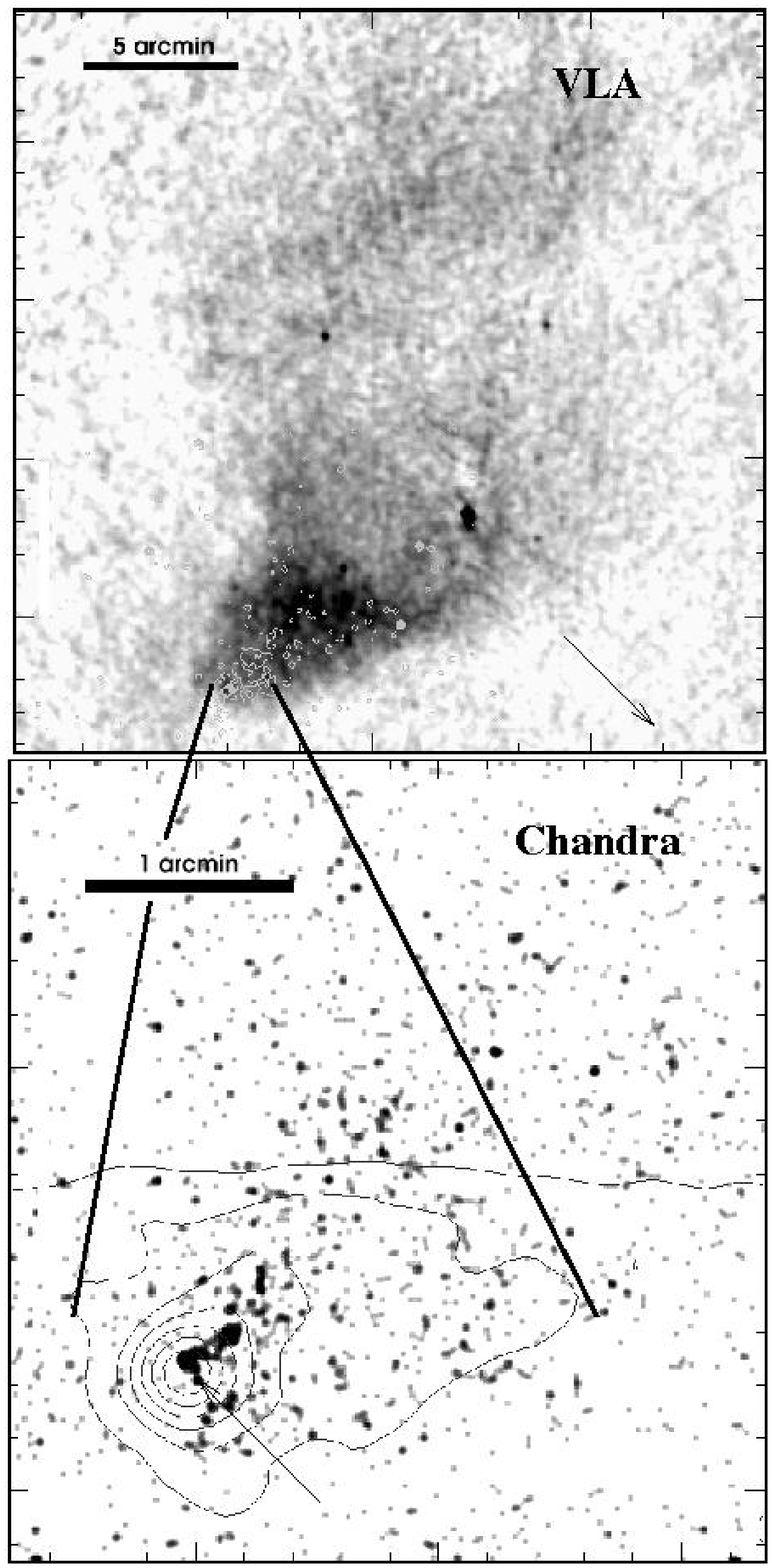, height=7cm}
\epsfig{file=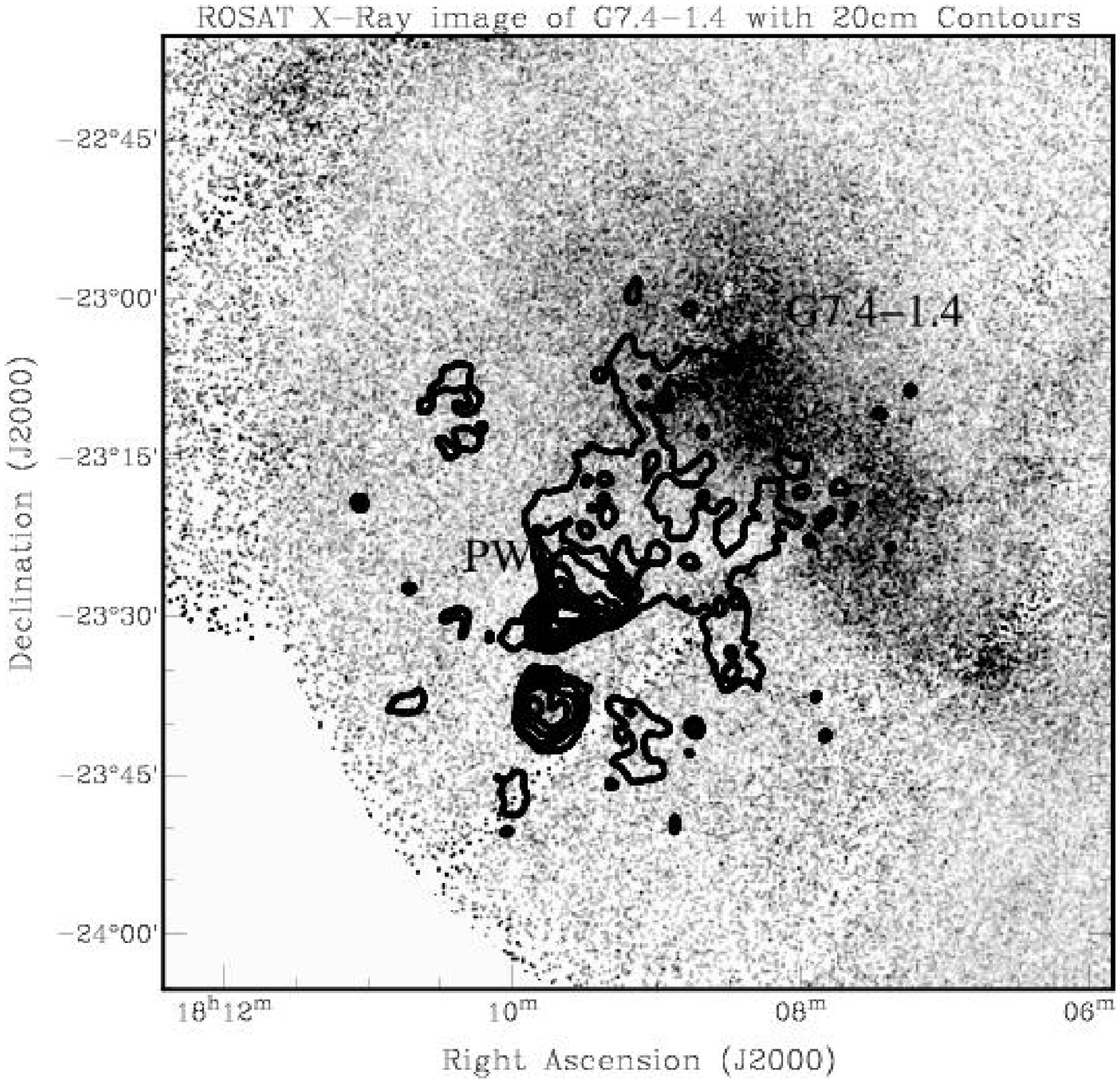, height=7cm}
\caption{{\it Left:} VLA and Chandra images of the PWN in GeV J1809$-$2327. 
The contours on the Chandra image are from the EPIC-PN, which are cut off
due to the small window used. {\it Right:} $ROSAT$ image of the SNR G7.4$-$1.4, 
with contours from the VLA image of the PWN.}
\label{1809_img}
\end{figure}

Remarkably, several RPWN have been discovered in the error boxes of
$EGRET$ sources which are apparently variable. The source with the
highest $V_{12}$ value in the Galactic plane without a potential Blazar
identification is GeV J1809$-$2327.  An X-ray nebulae was discovered 
in its error box with $ASCA$ \cite{rrk01} which subsequent Chandra and VLA
imaging showed to be a RPWN \cite{brrk02}. An XMM-Newton EPIC-PN 
observation, which was unfortunately taken in small window mode, suggests
a fainter X-ray halo surrounding the X-ray trail seen in the short 
Chandra exposure, strongly reminiscent of the Mouse (Fig.~\ref{1809_img}).
The X-ray emission
seen in the EPIC-PN image is coincident with the front edge of the radio 
PWN, itself reminiscent of the head of the Mouse, with the suggestion of 
a fainter, extended body to the Northwest. The radio and X-ray emission
seems to extend back towards a possible X-ray SNR, G7.4$-$1.4, 
discovered by $ROSAT$. Wide-field 90cm imaging is underway to determine if there
is a corresponding radio SNR. The PWN appears to be embedded in the 
Lynds 227 dark Nebula, and CO imaging suggests a physical 
connection between the PWN and the molecular cloud \cite{okn+99}.

The region near Galactic longitude $l=18^{\circ}$ contains several 
$EGRET$ sources, one or more of which are probably associated with 
the well known unidentified $1-30$ MeV Comptel source 
(see Collmar et al., these proceedings). 
Wide-field radio imaging 
(Brogan et al. in preparation) of the region shows several SNR and the Sharpless 53 cluster 
of H{\tt II} regions. 
One of the $EGRET$ sources, GeV J1825$-$1310, 
has the second highest $V_{12}$ value of Galactic plane sources. 
$ASCA$ imaging of this source followed by a short Chandra image again
revealed a faint RPWN in a stellar cluster (Romani et al. in preparation). 
There is some hint of a larger radio nebula in the 90cm image.  

Two other sources with $V_{12} > 1.3 \,(95\%$ confidence of variability)
contain apparent RPWN. One is the PWN around PSR B1853+01 in the X-ray 
composite SNR W44 \cite{pks02}. The other is the source containing the 
Kookaburra radio complex, the wings of which contain the PWN around PSR 
J1420$-$6048 and the Rabbit PWN \cite{rrjg99}. 
Chandra and XMM-Newton imaging of the 
Rabbit nebula (Ng et al., Roberts et al. in preparation) 
show the X-ray emission  going from a point source in the 
front paw back towards the body and the southern wing of the Kookaburra. 
This strongly suggests the Rabbit is the head of a RPWN, with the 
wing being the fainter body. In addition to these sources, there is a 
probable RPWN, G369.89$-$0.08 \cite{lwl03},
within the error box of the variable source near the Galactic center, 
which the Mouse is very near as well.

Overall, around half of the Galactic plane sources with good evidence
of variability and no candidate Blazar counterpart seem to contain 
RPWN. All of them also seem to be in regions containing molecular 
clouds or other evidence of relatively high densities in the ambient medium. 
This suggests the $\gamma$-ray emission may be associated with the
pulsar passing through local density enhancements.  

\vspace{-0.3cm}
\section{Conclusions}

\vspace{-0.3cm}
There is an intimate relationship between the generation of PWN and 
of $\gamma$-ray emission. This is not surprising since both indicate
particles being accelerated to very high energies in the pulsar
magnetosphere. Because of this, $\gamma$-ray sources in the Galaxy 
tend to pick out PWN that otherwise might be hard to find in
the crowded Galactic plane. SPWN with clear toroidal morphologies
are key to determining pulsar geometries, but require deep Chandra imaging.
No other X-ray telescope now operating or currently planned has the 
necessary resolution. RPWN appear to be associated with variable
$EGRET$ sources, which $AGILE$ and $GLAST$ should easily confirm. 
However, there is no real theory developed yet as to how they might be 
generating $\gamma$-ray emission, and much deeper Chandra X-ray 
imaging is required to determine their true structure. 

\vspace{-0.3cm}
\bibliographystyle{klunamed}

\bibliography{journals_apj,malloryrefs,modrefs,psrrefs,crossrefs}

\begin{thebibliography}{}

\bibitem[\protect\citeauthoryear{{Braje} et~al.}{2002}]{brrk02}
{Braje}, T.~M., R.~W. {Romani}, M.~S.~E. {Roberts}, and N. {Kawai}: 2002,
  `{Chandra Imaging of the Gamma-Ray Source GeV J1809-2327}'.
\newblock {\em ApJ} {\bf 565}, L91--L95.

\bibitem[\protect\citeauthoryear{{Gaensler}}{2004}]{gae04}
{Gaensler}, B.~M.: 2004, `{Shocks and Wind Bubbles Around Energetic Pulsars}'.
\newblock In: {\em IAU Symposium}. pp. 151--+.

\bibitem[\protect\citeauthoryear{{Gaensler} et~al.}{2003}]{gvc+03}
{Gaensler}, B.~M., E. {van der Swaluw}, F. {Camilo}, V.~M. {Kaspi}, F.~K.
  {Baganoff}, F. {Yusef-Zadeh}, and R.~N. {Manchester}: 2003, `{The Mouse That
  Soared: High Resolution X-ray Imaging of the Pulsar-Powered Bow Shock
  G359.23-0.82}'.
\newblock {\em ArXiv Astrophysics e-prints}.

\bibitem[\protect\citeauthoryear{{Halpern} et~al.}{2001}]{hcg+01}
{Halpern}, J.~P., F. {Camilo}, E.~V. {Gotthelf}, D.~J. {Helfand}, M. {Kramer},
  A.~G. {Lyne}, K.~M. {Leighly}, and M. {Eracleous}: 2001, `{PSR J2229+6114:
  Discovery of an Energetic Young Pulsar in the Error Box of the EGRET Source
  3EG J2227+6122}'.
\newblock {\em ApJ} {\bf 552}, L125--L128.

\bibitem[\protect\citeauthoryear{{Hessels} et~al.}{2004}]{hrr+04}
{Hessels}, J.~W.~T., M.~S.~E. {Roberts}, S.~M. {Ransom}, V.~M. {Kaspi}, R.~W.
  {Romani}, C.~. {Ng}, P.~C.~C. {Freire}, and B.~M. {Gaensler}: 2004,
  `{Observations of PSR J2021+3651 and its X-ray Pulsar Wind Nebula
  G75.2+0.1}'.
\newblock {\em ApJ} {\bf 612}, 389--397.

\bibitem[\protect\citeauthoryear{{Kaspi} et~al.}{2004}]{krh04}
{Kaspi}, V.~M., M.~S.~E. {Roberts}, and A.~K. {Harding}: 2004, `{Isolated
  Neutron Stars}'.
\newblock {\em ArXiv Astrophysics e-prints}.

\bibitem[\protect\citeauthoryear{Lamb and Macomb}{1997}]{lm97}
Lamb, R.~C. and D.~J. Macomb: 1997, `Point Sources of GeV Gamma Rays'.
\newblock {\em ApJ} {\bf 488}, 872.

\bibitem[\protect\citeauthoryear{{Lu} et~al.}{2003}]{lwl03}
{Lu}, F.~J., Q.~D. {Wang}, and C.~C. {Lang}: 2003, `{The Chandra Detection of
  Galactic Center X-Ray Features G359.89-0.08 and G359.54+0.18}'.
\newblock {\em AJ} {\bf 126}, 319--326.

\bibitem[\protect\citeauthoryear{{Ng} and {Romani}}{2004}]{nr04}
{Ng}, C.-Y. and R.~W. {Romani}: 2004, `{Fitting Pulsar Wind Tori}'.
\newblock {\em ApJ} {\bf 601}, 479--484.

\bibitem[\protect\citeauthoryear{{Nolan} et~al.}{2003}]{ntgm03}
{Nolan}, P.~L., W.~F. {Tompkins}, I.~A. {Grenier}, and P.~F. {Michelson}: 2003,
  `{Variability of EGRET Gamma-Ray Sources}'.
\newblock {\em ApJ} {\bf 597}, 615--627.

\bibitem[\protect\citeauthoryear{{Oka} et~al.}{1999}]{okn+99}
{Oka}, T., N. {Kawai}, T. {Naito}, T. {Horiuchi}, M. {Namiki}, Y. {Saito},
  R.~W. {Romani}, and T. {Kifune}: 1999, `A Dark Cloud Associated with an
  Unidentified EGRET Source'.
\newblock {\em ApJ} {\bf 526}, 764--771.

\bibitem[\protect\citeauthoryear{{Petre} et~al.}{2002}]{pks02}
{Petre}, R., K.~D. {Kuntz}, and R.~L. {Shelton}: 2002, `{The X-Ray Structure
  and Spectrum of the Pulsar Wind Nebula Surrounding PSR B1853+01 in W44}'.
\newblock {\em ApJ} {\bf 579}, 404--410.

\bibitem[\protect\citeauthoryear{Radhakrishnan and Cooke}{1969}]{rc69a}
Radhakrishnan, V. and D.~J. Cooke: 1969, `Magnetic poles and the polarization
  structure of pulsar radiation'.
\newblock {\em Astrophys. Lett.} {\bf 3}, 225--229.

\bibitem[\protect\citeauthoryear{{Roberts} et~al.}{2002}]{rhr+02}
{Roberts}, M.~S.~E., J.~W.~T. {Hessels}, S.~M. {Ransom}, V.~M. {Kaspi},
  P.~C.~C. {Freire}, F. {Crawford}, and D.~R. {Lorimer}: 2002, `{PSR
  J2021+3651: A Young Radio Pulsar Coincident with an Unidentified EGRET
  {$\gamma$}-Ray Source}'.
\newblock {\em ApJ} {\bf 577}, L19--L22.

\bibitem[\protect\citeauthoryear{Roberts et~al.}{1999}]{rrjg99}
Roberts, M. S.~E., R.~W. Romani, S. Johnston, and A.~J. Green: 1999, `The
  "Rabbit": A Potential Radio Counterpart of GeV J1417$-$6100'.
\newblock {\em ApJ} {\bf 515}, 712--720.

\bibitem[\protect\citeauthoryear{Roberts et~al.}{2001}]{rrk01}
Roberts, M. S.~E., R.~W. Romani, and N. Kawai: 2001, `The ASCA Catalog of
  Potential X-Ray Counterparts of GeV Sources'.
\newblock {\em ApJS} {\bf 133}, 451--465.

\end{thebibliography}

\end{article}
\end{document}